\documentclass[floatfix,superscriptaddress,twocolumn,showpacs,prl,amsmath,amssymb]{revtex4} 
\usepackage{graphicx} 
\usepackage[dvipdfm]{color}
\usepackage{amsmath}
\usepackage{txfonts}

\newcommand{\sub}[1]{$_{\mathrm {#1}}$}
\newcommand{\subm}[1]{_{\mathrm {#1}}}

\newcommand{\spsm}[1]{^{\mathrm {#1}}}
\newcommand{\etal}{\textit{et~al.}}

\renewcommand{\deg}{^{\circ}}
\newcommand{\Tc}{T\subm{c}}
\newcommand{\Tco}{T\subm{c}\spsm{onset}}

\newcommand{\x}{{\boldmath$X$}}

\newcommand{\Hcc}{H\subm{c2}}

\newcommand{\dash}{^{\prime}}
\newcommand{\tm}[1]{(TMTSF)\sub{2}{#1}}
\newcommand{\tmc}{\tm{ClO\sub{4}}}
\newcommand{\tmp}{\tm{PF\sub{6}}}
\newcommand{\tmx}{\tm{$X$}}
\newcommand{\cstar}{c^{\ast}}
\newcommand{\Rzz}{R_{c^{\ast}}}
\newcommand{\Hp}{H\subm{P}}

\newcommand{\qFFLO}{{\boldmath$q$}\sub{FFLO}}

\begin{document}

\title{Anomalous In-Plane Anisotropy of the Onset of Superconductivity in (TMTSF)$_2$ClO$_4$}

\author{Shingo~Yonezawa}
\affiliation{Department of Physics, Graduate School of Science, 
Kyoto University, Kyoto 606-8502, Japan}
\author{S.~Kusaba}
\affiliation{Department of Physics, Graduate School of Science, 
Kyoto University, Kyoto 606-8502, Japan}
\author{Y.~Maeno}
\affiliation{Department of Physics, Graduate School of Science, 
Kyoto University, Kyoto 606-8502, Japan}

\author{P.~Auban-Senzier}
\affiliation{Laboratoire de Physique des Solides (UMR 8502) - Universit{\'{e}} Paris-Sud, 91405 Orsay, France}
\author{C.~Pasquier}
\affiliation{Laboratoire de Physique des Solides (UMR 8502) - Universit{\'{e}} Paris-Sud, 91405 Orsay, France}

\author{K.~Bechgaard}
\affiliation{Department~of~Chemistry, Oersted~Institute, Universitetsparken 5, 2100 Copenhagen, Denmark}

\author{D.~J{\'{e}}rome}
\affiliation{Laboratoire de Physique des Solides (UMR 8502) - Universit{\'{e}} Paris-Sud, 91405 Orsay, France}

\email{yonezawa@scphys.kyoto-u.ac.jp}

\date{\today}

\begin{abstract}
We report the magnetic field-amplitude and field-angle dependence of 
the superconducting onset temperature $\Tco$ of the organic superconductor \tmc\ 
in magnetic fields {\boldmath$H$} accurately aligned to the conductive $ab\dash$ plane. 
We revealed that the rapid increase of the onset fields at low temperatures occurs 
both for {\boldmath$H$}${}\parallel b\dash$ and {\boldmath$H$}${}\parallel a$, 
irrespective of the carrier confinement. 
Moreover, in the vicinity of the Pauli limiting field, 
we report a shift of a principal axis of the in-plane field-angle dependence of $\Tco$ 
away from the $b\dash$ axis. 
This feature may be related to an occurrence of Fulde-Ferrell-Larkin-Ovchinnikov phases.
\end{abstract}

\maketitle

Since the discovery of the organic superconductors \tmx\ 
(where TMTSF stands for tetramethyl-tetraselena-fulvalene, 
$X=$ ClO\sub{4}, PF\sub{6}, etc.)~\cite{Jerome1980,Bechgaard1981},
their superconductivity has been studied with much attention.
Because of the strong anisotropy in the electrical conductivity of these materials~\cite{IshiguroYamajiText}, 
they provide excellent opportunities to study the properties of quasi-one-dimensional (Q1D)
superconductors.
One of the most important and controversial issues on the superconductivity of these materials 
is their superconducting (SC) pairing symmetry~\cite{Zhang2007AdvPhys}.
In this Letter, we provide experimental results
that contain new crucial clues in understanding the SC symmetry of \tmc.

It has been suggested that the superconductivity of \tmx\ is unconventional
with line(s) of node on its SC gap,
indicated through the NMR relaxation time~\cite{Takigawa1987}
and the impurity concentration dependence of the transition temperature $\Tc$~\cite{Joo2005}.
However, its SC symmetry is still controversial as we will review below.
One key feature of the SC symmetry is their unusually-high upper critical fields $\Hcc(T)$.
Lee \etal~\cite{Lee1997} reported that $\Hcc(T)$ of \tmp\ determined from resistivity
diverges as temperature decreases
and $\Hcc(T)$ reaches up to 80~kOe at the lowest temperatures
when magnetic fields {\boldmath$H$} are applied parallel to the $b\dash$ axis
(perpendicular to the most conductive $a$ axis in the $ab$ plane).
In this field direction, carriers are confined in the $ab$ plane
due to the field-induced dimensional crossover (FIDC) as the field increases~\cite{Strong1994,Joo2006}.
This confinement can be interpreted as 
a consequence of cyclotron motion of carriers on the Fermi surface (FS)~\cite{Zhang2007AdvPhys}.
The FIDC suppresses the orbital pair-breaking effect and 
may allow the superconductivity to survive in higher fields.
Interestingly,
80~kOe for $\Hcc\parallel b\dash$ far exceeds the so-called Pauli-Clogston limit $\Hp$~\cite{Clogston1962},
which fulfills a relation $\Hp/\Tc=18.4$~kOe/K for an isotropic gap,
where singlet Cooper pairs are unstable because unpaired carriers have a lower energy 
due to the Zeeman effect.
In the case of Ref.~\cite{Lee1997}, $\Hp$ was estimated to be 20~kOe.
Similar results have been obtained in \tmc\ 
by resistivity and magnetic torque measurements~\cite{Oh2004}.
One interpretation attributes this survival of superconductivity above $\Hp$ 
to a spin-triplet state~\cite{Lebed1999,Lebed2000}.
On the other hand, it has been pointed out that, in Q1D superconductors, 
even singlet superconductivity can be stable far above $\Hp$ by forming a spatially-modulating
SC state~\cite{Lebed1986,Dupuis1994,Miyazaki1999}, 
which is equivalent to the so-called Fulde-Ferrell-Larkin-Ovchinnikov (FFLO) state~\cite{Fulde1964,Larkin1965}.
In 2002, Lee \etal~\cite{Lee2001} reported 
the absence of a change in the ${}^{77}$Se Knight shift of \tmp\ at $\Tc$ under pressure,
in favor of a triplet scenario.
However, recently Shinagawa \etal~\cite{Shinagawa2007} 
observed a clear change of the ${}^{77}$Se Knight shift of \tmc\ at $\Tc$ in lower fields.
This finding motivated us to reexamine the possibility of singlet pairing in \tmx.

To resolve this puzzle, we are interested in the superconductivity in {\boldmath$H$}${}\parallel a$
and its in-plane anisotropy.
Although not much attention has been paid to 
the superconductivity for {\boldmath$H$}${}\parallel a$ so far,
data for $\Hcc(T)\parallel a$ of \tmp~\cite{Lee1997} looks quite interesting:
It has a steep slope near $H=0$,
but saturates when it reaches $\Hp$
probably due to the Pauli effect,
and it slightly increases again below 0.3~K.
However, for \tmc\ $\Hcc(T)\parallel a$ was reported only above 0.5~K~\cite{Murata1987}.
The in-plane anisotropy of $\Hcc$ of \tmc\ was also reported but only at 1.03~K~\cite{Murata1987},
where $\Hcc(T)$ is far below $\Hp$.

In the present study, we revealed the rapid increase of the onset fields
not only for {\boldmath$H$}${}\parallel b\dash$, where the electronic state becomes essentially 
2D due to the FIDC, 
but also for {\boldmath$H$}${}\parallel a$,
where the electronic state remains anisotropic 3D.
We also observed new features of the in-plane anisotropy developing above 20~kOe, 
which provide a crucial step to understand the origins of the enhancement of $\Hcc$,
in terms of FFLO states.


We used single crystals of \tmc\ grown by an electro-crystallization technique,
with dimensions of approximately $2.0\times 0.2\times 0.1$~mm$^3$.
We report here the results of the sample with the highest $\Tc$ among up to 10 samples.
We note that we obtained similar results in another sample.
The resistance along the $\cstar$ axis $\Rzz$ was measured using an ac four-probe method.
(The direction of the $\cstar$ axis is perpendicular to the $ab\dash$ plane
and is the least conductive direction.)
The measurements were performed with a dilution refrigerator down to 80~mK.
Temperature was measured using a RuO$_2$ resistance thermometer
with magnetoresistance correction.
The anion ordering temperature of \tmc\ is 24~K.
Therefore, in the temperature interval between 25~K and 22~K,
a cooling rate as slow as 2~mK/min was chosen 
to ensure that all anions are ordered and the whole sample is in the ``relaxed state''.

\begin{figure}
\includegraphics[width=7.5cm]{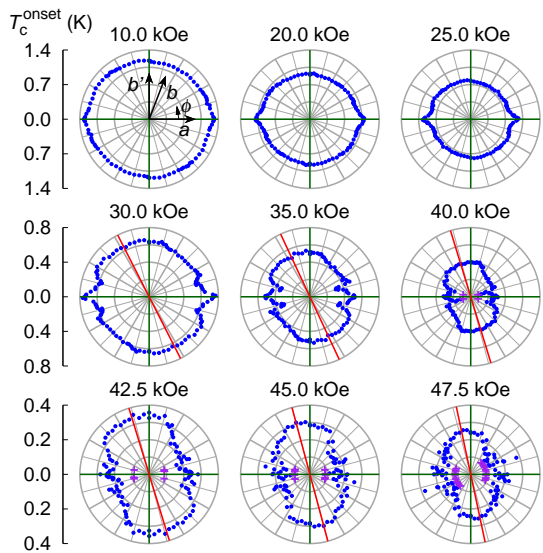}
\caption{(color online)
Polar plots of $\Tco(\phi)$.
The points for $|\phi|>90\deg$ are from the same data as those for $|\phi|\le 90\deg$.
They are plotted in order to symmetrize the figure.
Error bars are omitted for the sake of clarity.
Purple crosses indicate that no SC anomaly in $\Delta\sigma(T)$ was observed above 80~mK.
The directions of some crystalline principal axes are illustrated in the upper-left graph.
Solid red lines indicate the direction of the new principal axis {\boldmath$X$}.
\label{fig:polar-plot}}
\end{figure}

Magnetic fields are applied using the ``Vector Magnet'' system~\cite{Deguchi2004RSI},
with which we can control the field direction without mechanical heatings.
The directions of the orthogonal crystalline axes (the $a$, $b\dash$, and $\cstar$ axes) 
of the sample were determined from the anisotropy of $\Hcc$ at 0.1~K.
The accuracy of field alignment with respect to the $ab\dash$ plane 
and of the $a$ axis within the $ab\dash$ plane are both better than 0.1~degree.
We also determined the directions of the triclinic crystalline axes (the $b$ and $c$ axes)
from angular magnetoresistance oscillations.
The details of these procedures will be presented elsewhere.
Hereafter, we denote 
the azimuthal angle within the $ab\dash$ plane as $\phi$
which is measured from the $a$ axis.
We defined $\phi$ so that the $b$ axis lies in the quadrant $0\deg<\phi<90\deg$
as indicated in Fig.~\ref{fig:polar-plot}.


\begin{figure}
\includegraphics[width=7.5cm]{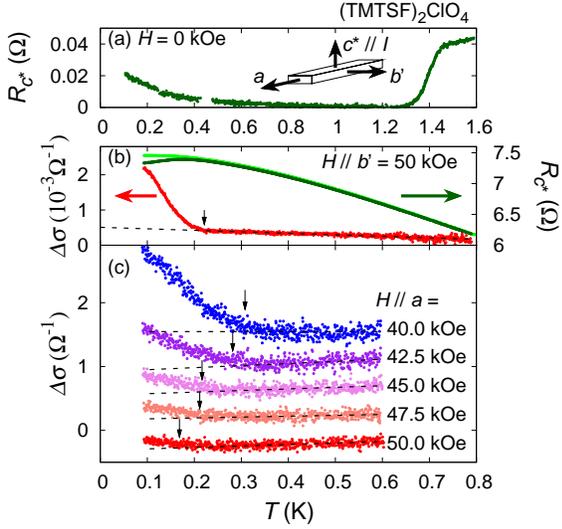}
\caption{(color online)
(a) Temperature dependence of $\Rzz$ in zero field.
The directions of the orthogonal crystalline axes are illustrated in the inset.
(b) Conductance difference $\Delta\sigma\equiv R_{\cstar}^{-1}(H_{\cstar}=0)-R_{\cstar}^{-1}(H_{\cstar}>0)$
under an in-plane 50-kOe field applied parallel to the $b\dash$ axis.
Resistances for $H_{\cstar}=0$~kOe and $H_{\cstar}=0.5$~kOe are plotted against the right vertical axis.
(c) Temperature dependence of $\Delta\sigma$ with the in-plane field 
applied parallel to the $a$ axis and $H_{\cstar}=1.0$~kOe.
Some data are shifted vertically for clarity.
\label{fig:example}}
\end{figure}

We first present $\Rzz(T)$ in zero field in Fig.~\ref{fig:example}(a).
Although $\Rzz$ of this sample started to drop at as high as 1.45~K and reached zero at 1.30~K, 
$\Rzz$ increases again below 0.8~K.
This increase, which is almost independent of magnetic fields, 
is probably attributed to small cracks in the sample.
The data of $\Rzz(T)$ for {\boldmath$H$}${}\parallel b\dash$ at 50~kOe
is presented in Fig.~\ref{fig:example}(b).
We observed a decrease of $\Rzz$ below 0.2~K, consistent with a previous report~\cite{Lee1995}.
In order to confirm that such a decrease is due to a superconducting contribution,
we measured $\Rzz(T)$ after adding a small out-of-plane component $H_{\cstar}={}$0.5-1.0~kOe to the magnetic field.
If this decrease is due to the superconductivity, $H_{\cstar}$ should suppress the superconductivity and 
eliminate the decrease of $\Rzz(T)$. 
As plotted in Fig.~\ref{fig:example}(b), the decrease was indeed eliminated by adding $H_{\cstar}$.
Therefore, it is confirmed that the decrease of $\Rzz(T)$ is a contribution of the superconductivity.
We used the following procedures to define the onset temperature of superconductivity $\Tco$:
We evaluated the conductance difference $\Delta\sigma\equiv \Rzz^{-1}(H_{\cstar}=0)-\Rzz^{-1}(H_{\cstar}>0)$
and defined $\Tco$ as the temperature at which $\Delta\sigma(T)$ exhibits a sharp increase,
as marked by the small arrow in Fig.~\ref{fig:example}(b).
This definition characterizes the very onset of superconductivity.
We note that this anomaly in $\Delta\sigma(T)$ is not due to the normal state magnetoresistance,
because it is unlikely that an abrupt change in the difference between $\Rzz(H_{\cstar}=0)$ and $\Rzz(H_{\cstar}>0)$
occurs at a certain temperature.
The definition has the advantage that $\Tco$ is not affected by the extrinsic small increase of $\Rzz$ 
because it is cancelled in the subtraction.
For {\boldmath$H$}${}\parallel \cstar$, $\Tco(H)$ was determined similarly from the conductance difference
$\Delta\sigma(H) \equiv \Rzz^{-1}(H_{\cstar}=H)\,-\,\Rzz^{-1}(H_{\cstar}=H+\Delta H)$.

\begin{figure}
\includegraphics[width=7.5cm]{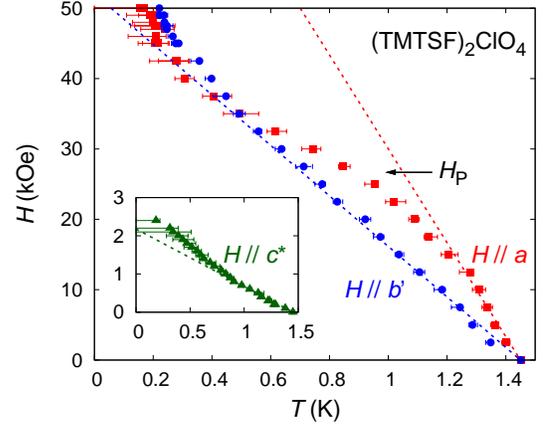}
\caption{(color online)
Magnetic fields vs. $\Tco$ of \tmc\ for {\boldmath$H$}${}\parallel a$ (filled squares) and 
{\boldmath$H$}${}\parallel b\dash$ (filled circles).
The arrow indicates the estimated value of $\Hp=26.7$~kOe.
The phase diagram for {\boldmath$H$}${}\parallel \cstar$ is shown in the inset.
The broken lines indicate the calculated initial slopes using transfer integrals 
explained in the text.
\label{fig:phase_diagram}}
\end{figure}

The phase diagrams for {\boldmath$H$}${}\parallel a$,
{\boldmath$H$}${}\parallel b\dash$, and {\boldmath$H$}${}\parallel \cstar$ 
are presented in Fig.~\ref{fig:phase_diagram}.
In the vicinity of $H=0$, linear temperature dependences of the curves were observed for all field directions,
which can be analyzed in a GL model for a clean type-II superconductor.
Within a GL theory with a tight-binding model, 
the slope $d\Hcc(T)/dT$ at $\Tc(H=0)$ is related to the transfer integral $t$ of each direction~\cite{Gorkov1985}.
By taking into account the $k_{z}$ dependence and the nodes of the gap over the FS, 
we obtain $t_a=1200$~K, $t_{b\dash}=310$~K, and $t_{\cstar}=7.0$~K
from the initial slopes indicated by the broken lines in Fig.~\ref{fig:phase_diagram}. 
These values agree favorably with realistic band parameters~\cite{IshiguroYamajiText}.
From these analysis, it is clear that $\Hcc(T)$ is governed by the orbital limitation
at low fields in all three directions.

In higher fields, the behavior of these curves is qualitatively different.
The curve for {\boldmath$H$}${}\parallel b\dash$ keeps a linear temperature dependence up to
35~kOe and starts to exhibit a rapid upturn in higher fields.
This behavior is consistent with the ``initial cool'' curve in Ref.~\cite{Oh2004}.
For {\boldmath$H$}${}\parallel a$, the curve apparently shows limiting behavior;
this is consistent with the Pauli-limiting behavior with the estimated value of $\Hp=26.7$~kOe.
Interestingly, in higher fields, a small kink of $\Delta\sigma$
remains visible up to 50~kOe also for {\boldmath$H$}${}\parallel a$, as shown in Fig.~\ref{fig:example}(c).
Consequently, the onset curve diverges for {\boldmath$H$}${}\parallel a$ at low temperatures, too.
This is, to our knowledge, the first report of the low-temperature 
high-field phase diagram of \tmc\ for {\boldmath$H$}${}\parallel a$.
We note that we obtained similar phase diagrams for another sample.
It is interesting that all three onset curves in Fig.~\ref{fig:phase_diagram} 
look similar to those for \tmp~\cite{Lee1997}.

Next, we focus on how $\Tco$ changes 
when magnetic fields are rotated in the $ab\dash$ plane.
The data are displayed in Fig.~\ref{fig:polar-plot} using polar plots of $\Tco(\phi)$,
where the direction of each point seen from the origin corresponds to the field direction and
the distance from the origin corresponds to $\Tco$.
At low fields, $\Tco(\phi)$ exhibits a sharp cusp at $\phi=0\deg$ ({\boldmath$H$}${}\parallel a$)
and a broad minimum around $\phi=\pm90\deg$ ({\boldmath$H$}${}\parallel b\dash$).
These low-field results are consistent with $\Hcc(\phi)$ reported by Murata \etal~\cite{Murata1987},
although the sharp peak at $\phi=0\deg$ cannot be explained in an anisotropic 3D GL theory~\cite{Huang1989}.
The chain-like crystal structure may play an important role in generating the sharp peak.

As the field increases above 20~kOe, another anomaly  
\textit{i.\,e.}\ dips of $\Tco(\phi)$, emerges at $|\phi|=\phi\subm{dip} = 17\pm 1\deg$.
We note that $\Rzz(T)$ in the normal state exhibits non-metallic temperature dependence 
for $|\phi|>\phi\subm{3D-2D}=19\pm 1\deg$ above 20~kOe,
signaling the onset of the FIDC~\cite{Strong1994,Joo2006}.
Because $\phi\subm{dip}\simeq \phi\subm{3D-2D}$, 
we infer that these dips are related to the FIDC.
When the dimensionality of the electronic system is lowered,
superconductivity in in-plane magnetic fields becomes more stable
because the orbital pair-breaking effect is suppressed.
Thus $\Tco(\phi)$ should be enhanced for $|\phi|>\phi\subm{3D-2D}$, 
resulting in a minimum of $\Tco(\phi)$ around $\phi\subm{3D-2D}$.
%

\begin{figure}
\includegraphics[width=7.5cm]{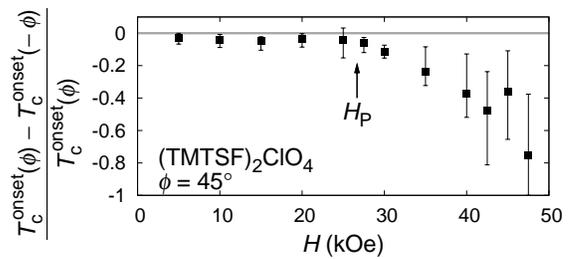}
\caption{Field dependence of the relative difference between $\Tco(+45\deg)$ and $\Tco(-45\deg)$.
Appearance of the new principal axis {\boldmath$X$} results in finite values of this quantity.
The arrow indicates the estimated value of $\Hp=26.7$~kOe.
\label{fig:asymmetry}}
\end{figure}

The third and probably the most important anomaly is, that in magnetic fields above 30~kOe, 
the $b\dash$ axis is no longer a symmetry axis of $\Tco(\phi)$ and 
a new principal axis {\boldmath$X$} appears around $\phi\sim-70\deg$ 
as indicated by the solid red lines in Fig.~\ref{fig:polar-plot}.
Moreover, behavior of $\Tco(\phi)$ around {\boldmath$X$},
a principal axis at high fields, and $b\dash$, a principal axis at low fields,
is qualitatively different:
At high fields $\Tco(\phi)$ is \textit{enhanced} around {\boldmath$X$},
while at low fields $\Tco(\phi)$ exhibits a broad local \textit{minimum} around the $b\dash$ axis.
In addition, this {\boldmath$X$} axis tends to rotate from $\phi\sim-70\deg$ toward 
the $b\dash$ axis as the field increases.
At the largest field available in this study, the deviation of {\boldmath$X$} from the $b\dash$ axis is 
reduced to about $10\deg$.
We checked that this change of symmetry is not due to a misalignment of the magnetic fields.
In Fig.~\ref{fig:asymmetry}, we plotted the relative difference between $\Tco(+45\deg)$ and $\Tco(-45\deg)$
against the field strength.
This quantity represents the asymmetry with respect to the $b\dash$ axis,
thus the appearance of {\boldmath$X$} results in finite values.
It is evident that the asymmetry, \textit{i.\,e.} {\boldmath$X$}, is absent in lower fields 
and then starts to develop around $\Hp$.
Therefore, the appearance of {\boldmath$X$} cannot be attributed to conventional origins 
like an anisotropy of the Fermi velocity, 
because variation of $\Tco(\phi)$ from such origins should develop from $H=0$.

We now discuss the origin of the new principal axis {\boldmath$X$}, 
which also indicates the direction of the field in which $\Tco(\phi)$ is enhanced. 
Its appearance should be related to the Pauli pair-breaking effect,
because {\boldmath$X$} appears at nearly $\Hp$. 
In the case of singlet pairing, the appearance of {\boldmath$X$} is attributable 
to the formation of an FFLO state~\cite{Fulde1964,Larkin1965},
in which the Cooper pairs have a finite wavevector \qFFLO.
In a Q1D superconductor, the stability of this state is greatly enhanced
by the nesting properties of its FS~\cite{Matsuda2007JPhysSocJpnReview}
and that \qFFLO\ essentially matches 
\emph{the nesting vector between the spin-up and the spin-down FSs},
which should be nearly parallel to the $a$ axis 
and should be independent of the field direction.

For {\boldmath$H$}${}\parallel b\dash$, it has been discussed using orthorhombic band structures that
an FFLO state with \qFFLO${}\parallel a$ 
becomes stable with a help of the FIDC~\cite{Lebed1986,Dupuis1994,Miyazaki1999}.
Although we are not aware of a publication on the in-plane field-angle variation of this FFLO state 
taking into account the realistic triclinic band structure of \tmc,
we expect that for $|\phi|>\phi\subm{3D-2D}$, where the FIDC takes place,
this FFLO state is still stable.
However, the direction of \qFFLO, matching with the nesting vector, should be slightly tilted from the $a$ axis
because of the triclinic FS of \tmc.
In addition, \qFFLO\ may vary with increasing the field 
because the separation between the spin-up and the spin-down FSs depends on $|${\boldmath$H$}$|$.
Within this scenario, one possible explanation of {\boldmath$X$}, 
which is the field-dependent special direction in the range $|\phi|>\phi\subm{3D-2D}$, 
is that \x\ is perpendicular to \qFFLO\ 
and thus {\boldmath$H$}${}\parallel{}${\boldmath$X$} corresponds to {\boldmath$H$}${}\perp{}$\qFFLO.
Because the direction of \qFFLO\ is expected to depend on the field strength as we explained,
{\boldmath$X$} ($\perp{}$\qFFLO) may also rotate in increasing field, 
which is consistent with our experimental results.

While for $|\phi|<\phi\subm{3D-2D}$, namely near {\boldmath$H$}${}\parallel a$,
the nature of superconductivity may differ from that near {\boldmath$H$}${}\parallel b\dash$
because of the absence of the FIDC.
Despite the absence of the FIDC,
the orbital-limiting field is much larger than $\Hp$ near {\boldmath$H$}${}\parallel a$ at low temperatures,
which is evident from the steep slope of $\Hcc(T)$ at $H=0$ in Fig.~\ref{fig:phase_diagram}.
An FFLO state in a Q1D system in fields parallel to the most conductive axis
has been proposed in a study of doped two-leg ladder cuprates 
using a $t$-$J$ model~\cite{Roux2006}. 
We infer that a similar FFLO state 
might be stable near {\boldmath$H$}${}\parallel a$,
although a theory adapted to \tmc, a coupled chain system, needs to be developed.
The recent NMR study, which reported that the density of states at the Fermi level 
recovers to the normal state value in the SC phase above 20~kOe for 
both {\boldmath$H$}${}\parallel a$ and {\boldmath$H$}${}\parallel b\dash$~\cite{Shinagawa2007},
would support these FFLO scenarios.

On the other hand, if \tmc\ is a triplet superconductor,
polarized Cooper pair spins may cause an anisotropy of $\Tco(\phi)$.
Assuming that the spins of the Cooper pairs are fixed to one direction,
superconductivity is not affected by a Pauli effect when the field is exactly parallel to the spins,
while it is suppressed for the other field directions.
In this case, however, it seems difficult to explain the rotation of {\boldmath$X$}.

In summary, we have studied the in-plane anisotropy of the superconducting onset temperature $\Tco$ of \tmc. 
We observed that $\Tco$ remains finite up to 50~kOe in magnetic fields parallel to the $a$ axis,
as well as for {\boldmath$H$}${}\parallel b\dash$. 
We suggest that the field-induced dimensional crossover plays an important role for 
the enhancement of $\Tco(\phi)$ when the field is tilted more than $17\deg$ from the $a$ axis.
In addition, we noticed that one of the principal axes for superconductivity,
which points along $b\dash$ at low fields, shifts away from this direction around 30~kOe 
but evolves back toward the $b\dash$ axis at higher fields.
The survival of superconductivity far above $\Hp$ and the unusual in-plane anisotropy
observed in the high field regime suggest the stabilization of modulated superconducting phases 
when high fields are aligned to the $ab\dash$ plane,
in favor of a spin-singlet scenario.
We speculate that two kinds of FFLO states are realized in this compound:
the one predicted by Dupuis \etal~\cite{Dupuis1994} near {\boldmath$H$}${}\parallel b\dash$
and the one related to the prediction by Roux \etal~\cite{Roux2006} for {\boldmath$H$}${}\parallel a$,
separated by the dips of $\Tco(\phi)$ around $\phi\sim \pm17\deg$.
We believe that theoretical studies taking into account the triclinic band structure are desirable
to understand our results and reveal the SC symmetry of \tmx.

We acknowledge Y. Machida, N. Joo, and M. Kriener for their supports.
We also acknowledge R. Ikeda, D. Poilblanc, G. Montambaux, and N. Dupuis for useful discussions.
This work has been supported by a Grant-in-Aid for the 21st Century COE 
``Center for Diversity and Universality in Physics'' from the Ministry of Education, Culture, Sports, Science and 
Technology (MEXT) of Japan.
It has also been supported by Grants-in-Aids for Scientific Research from MEXT 
and from the Japan Society for the Promotion of Science (JSPS).
S. Y. is financially supported as a JSPS Research Fellow.

\bibliography{../TMTSF}

\begin{thebibliography}{27}
\expandafter\ifx\csname natexlab\endcsname\relax\def\natexlab#1{#1}\fi
\expandafter\ifx\csname bibnamefont\endcsname\relax
  \def\bibnamefont#1{#1}\fi
\expandafter\ifx\csname bibfnamefont\endcsname\relax
  \def\bibfnamefont#1{#1}\fi
\expandafter\ifx\csname citenamefont\endcsname\relax
  \def\citenamefont#1{#1}\fi
\expandafter\ifx\csname url\endcsname\relax
  \def\url#1{\texttt{#1}}\fi
\expandafter\ifx\csname urlprefix\endcsname\relax\def\urlprefix{URL }\fi
\providecommand{\bibinfo}[2]{#2}
\providecommand{\eprint}[2][]{\url{#2}}

\bibitem[{\citenamefont{J{\'{e}}rome et~al.}(1980)\citenamefont{J{\'{e}}rome,
  Mazaud, Ribault, and Bechgaard}}]{Jerome1980}
\bibinfo{author}{\bibfnamefont{D.}~\bibnamefont{J{\'{e}}rome}},
  \bibinfo{author}{\bibfnamefont{A.}~\bibnamefont{Mazaud}},
  \bibinfo{author}{\bibfnamefont{M.}~\bibnamefont{Ribault}}, \bibnamefont{and}
  \bibinfo{author}{\bibfnamefont{K.}~\bibnamefont{Bechgaard}},
  \bibinfo{journal}{J. Phys. Lett.} \textbf{\bibinfo{volume}{41}},
  \bibinfo{pages}{95} (\bibinfo{year}{1980}).

\bibitem[{\citenamefont{Bechgaard et~al.}(1981)\citenamefont{Bechgaard,
  Carneiro, Olsen, Rasmussen, and Jacobsen}}]{Bechgaard1981}
\bibinfo{author}{\bibfnamefont{K.}~\bibnamefont{Bechgaard}},
  \bibinfo{author}{\bibfnamefont{K.}~\bibnamefont{Carneiro}},
  \bibinfo{author}{\bibfnamefont{M.}~\bibnamefont{Olsen}},
  \bibinfo{author}{\bibfnamefont{F.~B.} \bibnamefont{Rasmussen}},
  \bibnamefont{and} \bibinfo{author}{\bibfnamefont{C.~S.}
  \bibnamefont{Jacobsen}}, \bibinfo{journal}{Phys. Rev. Lett.}
  \textbf{\bibinfo{volume}{46}}, \bibinfo{pages}{852} (\bibinfo{year}{1981}).

\bibitem[{\citenamefont{Ishiguro et~al.}(1998)\citenamefont{Ishiguro, Yamaji,
  and Saito}}]{IshiguroYamajiText}
\bibinfo{author}{\bibfnamefont{T.}~\bibnamefont{Ishiguro}},
  \bibinfo{author}{\bibfnamefont{K.}~\bibnamefont{Yamaji}}, \bibnamefont{and}
  \bibinfo{author}{\bibfnamefont{G.}~\bibnamefont{Saito}},
  \emph{\bibinfo{title}{Organic Superconductors Second Edition}}
  (\bibinfo{publisher}{Springer-Verlag, Heidelberg}, \bibinfo{year}{1998}).

\bibitem[{\citenamefont{Zhang and {S\'{a} de Melo}}(2007)}]{Zhang2007AdvPhys}
\bibinfo{author}{\bibfnamefont{W.}~\bibnamefont{Zhang}} \bibnamefont{and}
  \bibinfo{author}{\bibfnamefont{C.~A.~R.} \bibnamefont{{S\'{a} de Melo}}},
  \bibinfo{journal}{Adv. Phys.} \textbf{\bibinfo{volume}{56}},
  \bibinfo{pages}{545} (\bibinfo{year}{2007}).

\bibitem[{\citenamefont{Takigawa et~al.}(1987)\citenamefont{Takigawa, Yasuoka,
  and Saito}}]{Takigawa1987}
\bibinfo{author}{\bibfnamefont{M.}~\bibnamefont{Takigawa}},
  \bibinfo{author}{\bibfnamefont{H.}~\bibnamefont{Yasuoka}}, \bibnamefont{and}
  \bibinfo{author}{\bibfnamefont{G.}~\bibnamefont{Saito}}, \bibinfo{journal}{J.
  Phys. Soc. Jpn.} \textbf{\bibinfo{volume}{56}}, \bibinfo{pages}{873}
  (\bibinfo{year}{1987}).

\bibitem[{\citenamefont{Joo et~al.}(2005)\citenamefont{Joo, Auban-Senzier,
  Pasquier, J{\'{e}}rome, and Bechgaard}}]{Joo2005}
\bibinfo{author}{\bibfnamefont{N.}~\bibnamefont{Joo}},
  \bibinfo{author}{\bibfnamefont{P.}~\bibnamefont{Auban-Senzier}},
  \bibinfo{author}{\bibfnamefont{C.~R.} \bibnamefont{Pasquier}},
  \bibinfo{author}{\bibfnamefont{D.}~\bibnamefont{J{\'{e}}rome}},
  \bibnamefont{and}
  \bibinfo{author}{\bibfnamefont{K.}~\bibnamefont{Bechgaard}},
  \bibinfo{journal}{Euro. Phys. Lett.} \textbf{\bibinfo{volume}{72}},
  \bibinfo{pages}{645} (\bibinfo{year}{2005}).

\bibitem[{\citenamefont{Lee et~al.}(1997)\citenamefont{Lee, Naughton, Danner,
  and Chaikin}}]{Lee1997}
\bibinfo{author}{\bibfnamefont{I.~J.} \bibnamefont{Lee}},
  \bibinfo{author}{\bibfnamefont{M.~J.} \bibnamefont{Naughton}},
  \bibinfo{author}{\bibfnamefont{G.~M.} \bibnamefont{Danner}},
  \bibnamefont{and} \bibinfo{author}{\bibfnamefont{P.~M.}
  \bibnamefont{Chaikin}}, \bibinfo{journal}{Phys. Rev. Lett.}
  \textbf{\bibinfo{volume}{78}}, \bibinfo{pages}{3555} (\bibinfo{year}{1997}).

\bibitem[{\citenamefont{Strong et~al.}(1994)\citenamefont{Strong, Clarke, and
  Anderson}}]{Strong1994}
\bibinfo{author}{\bibfnamefont{S.~P.} \bibnamefont{Strong}},
  \bibinfo{author}{\bibfnamefont{D.~G.} \bibnamefont{Clarke}},
  \bibnamefont{and} \bibinfo{author}{\bibfnamefont{P.~W.}
  \bibnamefont{Anderson}}, \bibinfo{journal}{Phys. Rev. Lett.}
  \textbf{\bibinfo{volume}{73}}, \bibinfo{pages}{1007} (\bibinfo{year}{1994}).

\bibitem[{\citenamefont{Joo et~al.}(2006)\citenamefont{Joo, Auban-Senzier,
  Pasquier, Yonezawa, Higashinaka, Maeno, Haddad, Charfi-Kaddour, Heritier,
  Bechgaard et~al.}}]{Joo2006}
\bibinfo{author}{\bibfnamefont{N.}~\bibnamefont{Joo}},
  \bibinfo{author}{\bibfnamefont{P.}~\bibnamefont{Auban-Senzier}},
  \bibinfo{author}{\bibfnamefont{C.~R.} \bibnamefont{Pasquier}},
  \bibinfo{author}{\bibfnamefont{S.}~\bibnamefont{Yonezawa}},
  \bibinfo{author}{\bibfnamefont{R.}~\bibnamefont{Higashinaka}},
  \bibinfo{author}{\bibfnamefont{Y.}~\bibnamefont{Maeno}},
  \bibinfo{author}{\bibfnamefont{S.}~\bibnamefont{Haddad}},
  \bibinfo{author}{\bibfnamefont{S.}~\bibnamefont{Charfi-Kaddour}},
  \bibinfo{author}{\bibfnamefont{M.}~\bibnamefont{Heritier}},
  \bibinfo{author}{\bibfnamefont{K.}~\bibnamefont{Bechgaard}},
  \bibnamefont{et~al.}, \bibinfo{journal}{Euro. Phys. J. B}
  \textbf{\bibinfo{volume}{52}}, \bibinfo{pages}{337} (\bibinfo{year}{2006}).

\bibitem[{\citenamefont{Clogston}(1962)}]{Clogston1962}
\bibinfo{author}{\bibfnamefont{A.~M.} \bibnamefont{Clogston}},
  \bibinfo{journal}{Phys. Rev. Lett.} \textbf{\bibinfo{volume}{9}},
  \bibinfo{pages}{266} (\bibinfo{year}{1962}).

\bibitem[{\citenamefont{Oh and Naughton}(2004)}]{Oh2004}
\bibinfo{author}{\bibfnamefont{J.~I.} \bibnamefont{Oh}} \bibnamefont{and}
  \bibinfo{author}{\bibfnamefont{M.~J.} \bibnamefont{Naughton}},
  \bibinfo{journal}{Phys. Rev. Lett.} \textbf{\bibinfo{volume}{92}},
  \bibinfo{pages}{067001} (\bibinfo{year}{2004}).

\bibitem[{\citenamefont{Lebed}(1999)}]{Lebed1999}
\bibinfo{author}{\bibfnamefont{A.~G.} \bibnamefont{Lebed}},
  \bibinfo{journal}{Phys. Rev. B} \textbf{\bibinfo{volume}{59}},
  \bibinfo{pages}{R721} (\bibinfo{year}{1999}).

\bibitem[{\citenamefont{Lebed et~al.}(2000)\citenamefont{Lebed, Machida, and
  Ozaki}}]{Lebed2000}
\bibinfo{author}{\bibfnamefont{A.~G.} \bibnamefont{Lebed}},
  \bibinfo{author}{\bibfnamefont{K.}~\bibnamefont{Machida}}, \bibnamefont{and}
  \bibinfo{author}{\bibfnamefont{M.}~\bibnamefont{Ozaki}},
  \bibinfo{journal}{Phys. Rev. B} \textbf{\bibinfo{volume}{62}},
  \bibinfo{pages}{R795} (\bibinfo{year}{2000}).

\bibitem[{\citenamefont{Lebed}(1986)}]{Lebed1986}
\bibinfo{author}{\bibfnamefont{A.~G.} \bibnamefont{Lebed}},
  \bibinfo{journal}{JETP Lett.} \textbf{\bibinfo{volume}{44}},
  \bibinfo{pages}{114} (\bibinfo{year}{1986}).

\bibitem[{\citenamefont{Dupuis and Montambaux}(1994)}]{Dupuis1994}
\bibinfo{author}{\bibfnamefont{N.}~\bibnamefont{Dupuis}} \bibnamefont{and}
  \bibinfo{author}{\bibfnamefont{G.}~\bibnamefont{Montambaux}},
  \bibinfo{journal}{Phys. Rev. B} \textbf{\bibinfo{volume}{49}},
  \bibinfo{pages}{8993} (\bibinfo{year}{1994}).

\bibitem[{\citenamefont{Miyazaki et~al.}(1999)\citenamefont{Miyazaki, Kishigi,
  and Hasegawa}}]{Miyazaki1999}
\bibinfo{author}{\bibfnamefont{M.}~\bibnamefont{Miyazaki}},
  \bibinfo{author}{\bibfnamefont{K.}~\bibnamefont{Kishigi}}, \bibnamefont{and}
  \bibinfo{author}{\bibfnamefont{Y.}~\bibnamefont{Hasegawa}},
  \bibinfo{journal}{J. Phys. Soc. Jpn.} \textbf{\bibinfo{volume}{68}},
  \bibinfo{pages}{3794} (\bibinfo{year}{1999}).

\bibitem[{\citenamefont{Fulde and Ferrell}(1964)}]{Fulde1964}
\bibinfo{author}{\bibfnamefont{P.}~\bibnamefont{Fulde}} \bibnamefont{and}
  \bibinfo{author}{\bibfnamefont{R.~A.} \bibnamefont{Ferrell}},
  \bibinfo{journal}{Phys. Rev.} \textbf{\bibinfo{volume}{135}},
  \bibinfo{pages}{A550} (\bibinfo{year}{1964}).

\bibitem[{\citenamefont{Larkin and Ovchinnikov}(1965)}]{Larkin1965}
\bibinfo{author}{\bibfnamefont{A.~I.} \bibnamefont{Larkin}} \bibnamefont{and}
  \bibinfo{author}{\bibfnamefont{Y.~N.} \bibnamefont{Ovchinnikov}},
  \bibinfo{journal}{Sov. Phys. JETP} \textbf{\bibinfo{volume}{20}},
  \bibinfo{pages}{762} (\bibinfo{year}{1965}).

\bibitem[{\citenamefont{Lee et~al.}(2001)\citenamefont{Lee, Brown, Clark,
  Strouse, Naughton, Kang, and Chaikin}}]{Lee2001}
\bibinfo{author}{\bibfnamefont{I.~J.} \bibnamefont{Lee}},
  \bibinfo{author}{\bibfnamefont{S.~E.} \bibnamefont{Brown}},
  \bibinfo{author}{\bibfnamefont{W.~G.} \bibnamefont{Clark}},
  \bibinfo{author}{\bibfnamefont{M.~J.} \bibnamefont{Strouse}},
  \bibinfo{author}{\bibfnamefont{M.~J.} \bibnamefont{Naughton}},
  \bibinfo{author}{\bibfnamefont{W.}~\bibnamefont{Kang}}, \bibnamefont{and}
  \bibinfo{author}{\bibfnamefont{P.~M.} \bibnamefont{Chaikin}},
  \bibinfo{journal}{Phys. Rev. Lett.} \textbf{\bibinfo{volume}{88}},
  \bibinfo{pages}{017004} (\bibinfo{year}{2001}).

\bibitem[{\citenamefont{Shinagawa et~al.}(2007)\citenamefont{Shinagawa,
  Kurosaki, Zhang, Parker, Brown, J{\'{e}}rome, Christensen, and
  Bechgaard}}]{Shinagawa2007}
\bibinfo{author}{\bibfnamefont{J.}~\bibnamefont{Shinagawa}},
  \bibinfo{author}{\bibfnamefont{Y.}~\bibnamefont{Kurosaki}},
  \bibinfo{author}{\bibfnamefont{F.}~\bibnamefont{Zhang}},
  \bibinfo{author}{\bibfnamefont{C.}~\bibnamefont{Parker}},
  \bibinfo{author}{\bibfnamefont{S.~E.} \bibnamefont{Brown}},
  \bibinfo{author}{\bibfnamefont{D.}~\bibnamefont{J{\'{e}}rome}},
  \bibinfo{author}{\bibfnamefont{J.~B.} \bibnamefont{Christensen}},
  \bibnamefont{and}
  \bibinfo{author}{\bibfnamefont{K.}~\bibnamefont{Bechgaard}},
  \bibinfo{journal}{Phys. Rev. Lett.} \textbf{\bibinfo{volume}{98}},
  \bibinfo{pages}{147002} (\bibinfo{year}{2007}).

\bibitem[{\citenamefont{Murata et~al.}(1987)\citenamefont{Murata, Tokumoto,
  Anzai, Kajimura, and Ishiguro}}]{Murata1987}
\bibinfo{author}{\bibfnamefont{K.}~\bibnamefont{Murata}},
  \bibinfo{author}{\bibfnamefont{M.}~\bibnamefont{Tokumoto}},
  \bibinfo{author}{\bibfnamefont{H.}~\bibnamefont{Anzai}},
  \bibinfo{author}{\bibfnamefont{K.}~\bibnamefont{Kajimura}}, \bibnamefont{and}
  \bibinfo{author}{\bibfnamefont{T.}~\bibnamefont{Ishiguro}},
  \bibinfo{journal}{Jpn. J. Appl. Phys.} \textbf{\bibinfo{volume}{26}},
  \bibinfo{pages}{1367} (\bibinfo{year}{1987}).

\bibitem[{\citenamefont{Deguchi et~al.}(2004)\citenamefont{Deguchi, Ishiguro,
  and Maeno}}]{Deguchi2004RSI}
\bibinfo{author}{\bibfnamefont{K.}~\bibnamefont{Deguchi}},
  \bibinfo{author}{\bibfnamefont{T.}~\bibnamefont{Ishiguro}}, \bibnamefont{and}
  \bibinfo{author}{\bibfnamefont{Y.}~\bibnamefont{Maeno}},
  \bibinfo{journal}{Rev. Sci. Instrum.} \textbf{\bibinfo{volume}{75}},
  \bibinfo{pages}{1188} (\bibinfo{year}{2004}).

\bibitem[{\citenamefont{Lee et~al.}(1995)\citenamefont{Lee, Hope, Leone, and
  Naughton}}]{Lee1995}
\bibinfo{author}{\bibfnamefont{I.~J.} \bibnamefont{Lee}},
  \bibinfo{author}{\bibfnamefont{A.~P.} \bibnamefont{Hope}},
  \bibinfo{author}{\bibfnamefont{M.~J.} \bibnamefont{Leone}}, \bibnamefont{and}
  \bibinfo{author}{\bibfnamefont{M.~J.} \bibnamefont{Naughton}},
  \bibinfo{journal}{Synth. Metals} \textbf{\bibinfo{volume}{70}},
  \bibinfo{pages}{747} (\bibinfo{year}{1995}).

\bibitem[{\citenamefont{Gor'kov and J{\'{e}}rome}(1985)}]{Gorkov1985}
\bibinfo{author}{\bibfnamefont{L.~P.} \bibnamefont{Gor'kov}} \bibnamefont{and}
  \bibinfo{author}{\bibfnamefont{D.}~\bibnamefont{J{\'{e}}rome}},
  \bibinfo{journal}{J. Phys. Lett.} \textbf{\bibinfo{volume}{46}},
  \bibinfo{pages}{L643} (\bibinfo{year}{1985}).

\bibitem[{\citenamefont{Huang and Maki}(1989)}]{Huang1989}
\bibinfo{author}{\bibfnamefont{X.}~\bibnamefont{Huang}} \bibnamefont{and}
  \bibinfo{author}{\bibfnamefont{K.}~\bibnamefont{Maki}},
  \bibinfo{journal}{Phys. Rev. B} \textbf{\bibinfo{volume}{39}},
  \bibinfo{pages}{6459} (\bibinfo{year}{1989}).

\bibitem[{\citenamefont{Matsuda and
  Shimahara}(2007)}]{Matsuda2007JPhysSocJpnReview}
\bibinfo{author}{\bibfnamefont{Y.}~\bibnamefont{Matsuda}} \bibnamefont{and}
  \bibinfo{author}{\bibfnamefont{H.}~\bibnamefont{Shimahara}},
  \bibinfo{journal}{J. Phys. Soc. Jpn.} \textbf{\bibinfo{volume}{76}},
  \bibinfo{pages}{051005} (\bibinfo{year}{2007}).

\bibitem[{\citenamefont{Roux et~al.}(2006)\citenamefont{Roux, White, Capponi,
  and Poilblanc}}]{Roux2006}
\bibinfo{author}{\bibfnamefont{G.}~\bibnamefont{Roux}},
  \bibinfo{author}{\bibfnamefont{S.~R.} \bibnamefont{White}},
  \bibinfo{author}{\bibfnamefont{S.}~\bibnamefont{Capponi}}, \bibnamefont{and}
  \bibinfo{author}{\bibfnamefont{D.}~\bibnamefont{Poilblanc}},
  \bibinfo{journal}{Phys. Rev. Lett.} \textbf{\bibinfo{volume}{97}},
  \bibinfo{pages}{087207} (\bibinfo{year}{2006}).

\end{thebibliography}

\end{document}